\documentclass[conference]{IEEEtran}

\usepackage{cite}
\usepackage[cmex10]{amsmath}
\usepackage{array}
\usepackage{mdwmath}
\usepackage{mdwtab}
\usepackage[tight,footnotesize]{subfigure}
\usepackage{tikz}

\usepackage[T1]{fontenc}
\usepackage{amssymb,amsfonts,amsthm}
\usepackage{graphicx,epsfig,psfrag,color}
\usepackage{times}
\usepackage{dsfont}
\usepackage{algorithm}
\usepackage[noend]{algorithmic}
\usepackage{caption}

\usepackage{array}
\usepackage{mdwmath}
\usepackage{mdwtab}
\usepackage{tikz}
\usepackage{graphicx,epsfig,psfrag,color}
\usepackage{multirow}

\usetikzlibrary{decorations.pathreplacing}

\tikzstyle{vertex}=[circle,fill=black!15,minimum size=20pt,inner sep=0pt,font=\footnotesize]
\tikzstyle{smallvertex}=[vertex,minimum size=10pt]
\tikzstyle{operator}=[vertex,fill=black!1]
\tikzstyle{smalloperator}=[circle,inner sep=0pt,minimum size=10pt,font=\footnotesize]
\tikzstyle{source} = [vertex, fill=red!34]
\tikzstyle{supersource} = [vertex, fill=blue!34,minimum size=15pt]
\tikzstyle{hiddensource} = [vertex, fill=red!8,minimum size=10pt]
\tikzstyle{smallsource} = [vertex, fill=red!34,minimum size=10pt]
\tikzstyle{receiver} = [vertex, fill=green!34]
\tikzstyle{smallreceiver} = [vertex, fill=green!34,minimum size=10pt]
\tikzstyle{edge} = [draw,thick,->]
\tikzstyle{undirect_edge} = [draw, thick]
\tikzstyle{dedge} = [edge,dotted]
\tikzstyle{redge} = [edge,color=red]
\tikzstyle{bedge} = [edge,color=blue]
\tikzstyle{gedge} = [edge,color=green]
\tikzstyle{medge} = [edge,color=magenta]
\tikzstyle{oedge} = [edge,color=orange]
\tikzstyle{bredge} = [edge,color=brown,line width=2pt]
\tikzstyle{weight} = [font=\footnotesize]
\tikzstyle{selected edge} = [draw,line width=5pt,-,red!50]

\newcommand{\bi}{\begin{itemize}}
\newcommand{\ei}{\end{itemize}}
\newcommand{\bal}{\begin{align}}
\newcommand{\eal}{\end{align}}

\newcommand{\thmref}[1]{Theorem~\ref{#1}}

\newcommand{\bG}{\mathbf{G}}
\newcommand{\bY}{\mathbf{Y}}
\newcommand{\bX}{\mathbf{X}}
\newcommand{\bZ}{\mathbf{Z}}

\newtheorem{theorem}{Theorem}
\newtheorem{lemma}{Lemma}
\newtheorem{corollary}{Corollary}
\newtheorem{definition}{Definition}
\newtheorem{remark}{Remark}
\newtheorem{const}{Coding Scheme}

\begin{document}

\title{Efficient Coding for Multi-source Networks using G\'acs-Korner Common Information}

\author{\IEEEauthorblockN{Salman Salamatian}
\IEEEauthorblockA{MIT,\\
Cambridge, MA, USA}
\and
\IEEEauthorblockN{Asaf Cohen}
\IEEEauthorblockA{Ben-Gurion University of the Negev\\
Beer-Sheva, 84105, Israel}
\and
\IEEEauthorblockN{Muriel M\'edard}
\IEEEauthorblockA{MIT,\\
Cambdridge, MA, USA}
}

\maketitle

\begin{abstract}

Consider a multi-source network coding problem with correlated sources. While the fundamental limits are known, achieving them, in general, involves a computational burden due to the complex decoding process. Efficient solutions, on the other hand, are by large based on source and network coding separation, thus imposing strict topological constraints on the networks which can be solved.

In this work, we introduce a novel notion of separation of source and network coding using G\'acs-K\"orner Common Information (CI). Unlike existing notions of separation, the sufficient condition for this separation to hold depends on the source structure rather than the network topology. Using the suggested separation scheme, we tackle three important multi-source problems. The first is the multi-source multicast. We construct efficient, zero error source codes, and via properties of the CI completely characterize the resulting rate region. The second is broadcast with side information. We establish a duality between this problem and the classical problem of degraded message set broadcast, and give two code constructions and their associated regions. Finally, we consider the Ahlswede-Korner problem in a network, and give an efficient solution which is tight under the CI constraints.
\end{abstract}

\begin{IEEEkeywords} Network Coding, Common Information, Distributed Source-Coding\end{IEEEkeywords}

\IEEEpeerreviewmaketitle

\section{Introduction}
Content distribution over a network is a topic of great interest in many practical areas, such as content delivery systems, design of proxies and mirror web-sites, cloud services, and so on. For the most part, these problems can be seen through the scope of a general multi-source network coding problem, which, since its inception by Ahlswede \emph{et al.} \cite{Ahlswede:2000th}, has been the subject of both theoretical and practical work. In fact, in many problem instances such as multi-source multicast with independent sources, there exist practical low-complexity schemes to achieve the fundamental limits \cite{Ho2006}. However, this is no longer true for more realistic settings, in particular, when the sources are correlated. In fact, while the fundamental limits of many of the problems we study herein are well-known, and usually follow from extensions of the pioneering works of Ahlswede \emph{et~al.} \cite{Ahlswede:2000th} and of Slepian and Wolf \cite{Slepian:1973wj}, no practical encoding and decoding algorithms fully address these problems in a multi-source network environment. Indeed, most present-day practical systems use very naive solutions to avoid redundancy, such as deduplication. 

Efficient schemes to solve these multi-source network coding problems have been the subject of some work. In particular, the problem of multi-source multicast. It has been shown that, in general, it is not possible to separate source and network coding 
\cite{Han:2010wa,Ramamoorthy:2006vr}, meaning that low-complexity implementations of Slepian-Wolf codes matched with low-complexity network-codes do not, in general, achieve the fundamental limits of communication, unless some topological conditions on the network are met. 
Joint source-network approaches have also been studied in \cite{Lee2007,Maierbacher:2009ug}, though usually such schemes suffer from either an implementation burden, or strong conditions on the network structure.

In this paper, we propose an alternate point of view on the problem, introducing new approaches based on the \emph{structure of the source correlation}, and, in particular, on the G\'acs-K\"orner Common Information between sources introduced in \cite{Gacs:1973vg}. This enables us to construct efficient codes for three problems of interest: Multi-source Multicast, Broadcast with Side Information, and Source Coding with Coded Side Information over networks.
The proposed constructions have low-complexity and require only zero-error lossless encoding, yielding a very tractable end-to-end solution. As the common information does not capture, in general, the correlation between sources entirely \cite{Gacs:1973vg}, our schemes will be restrictive in terms of the joint distribution of the sources. Nevertheless, we believe this setting is of interest in many practical cases. For example, in cloud storage where servers may hold either identical copies of files or a few versions, differing mainly by insertions, deletions and simple edits. Similarly, in sensor networks, where sensors may share a common state besides the additional data each sensor has. In these examples, the structure of the correlation is such that the common information is non-zero, and, in fact, can be very large. 

To compare our scheme with existing literature, we introduce three complexity levels, that range from exponential complexity algorithms such as joint-typicality, to polynomial time algorithms such as usual lossless source coding. We also make a distinction between asymptotically lossless, and zero-error code constructions. Thus, we say a code is \textbf{high-complexity}, if it requires an exponential complexity encoding or decoding as a function of the block length. Joint typicality, minimum entropy, or maximum likelihood algorithms fall into this category. A code has an \textbf{error-correcting complexity} (EC low-complexity), if it is a polynomial complexity code construction which succeeds asymptotically in terms of the block length. Efficient error-correcting codes and Wyner-schemes \cite{wyner74recent} based on those codes fall into this category. Finally, we say a code a \textbf{zero-error low-complexity}, if it is a polynomial zero-error code. Traditional prefix-free source coding, such as Huffman or Lempel-Ziv codes fall into this category.
\subsection{Main Contribution}
Our main results, and a comparison with the existing works are given in Table~\ref{tab:comp}.
Specifically, we first suggest a Source-Network separation scheme based on source decomposition. This allows us to effectively use very efficient point-to-point lossless source codes, without the complexity associated with joint typicality decoding or even with simpler distributed source codes, and with a zero-error guarantee. Using source decomposition inequalities, we completely characterize the rate region achievable with the suggested scheme, and discuss a few interesting special cases. Via a generalization of common information, we extend the results to any number of sources. In a sense, the above results replace the \emph{topological constraints} imposed on the network in order to have efficient separation, with \emph{source structure constraint}.

We then turn to the related problem of broadcast with side information. While the rate region is completely characterized under high-complexity decoding, in general, efficient distributed source codes cannot be applied without separation. We tackle this problem by proving a duality result, connecting this problem to that of broadcast with a \emph{degraded message set}, an allegedly unrelated problem which is open in general. Via this duality, we suggest two efficient constructions for the problem. The first is based on Polar codes, and assumes the network can support a degraded message set, while the other requires higher rates to the terminals, yet allows for the much simpler, source decomposition approach. 

Finally, we focus on source coding with a helper, a problem unsolved if the helper is located at an arbitrary location in the network. Instead of the strict topological constraints required for tight results in this case, we offer a separation scheme based on source decomposition.
\begin{table*}
\renewcommand{\arraystretch}{1.4}
{\normalsize
\centering
\begin{tabular}{l|c|c|c|}
\cline{2-4}
& High-Complexity & EC Low-Complexity & Zero-Error Low-Complexity \\ \hline
\multicolumn{1}{|c|}{ Network multisource multicast} & \cite{Song:2001tf}, \cite{ho2004network} & \cite{Han:2010wa}\IEEEauthorrefmark{2}, \cite{Ramamoorthy:2006vr}\IEEEauthorrefmark{2}, \cite{Salamatian2015}\IEEEauthorrefmark{2}, \cite{Lee2007}\IEEEauthorrefmark{2} \cite{Maierbacher:2009ug} \IEEEauthorrefmark{1}\IEEEauthorrefmark{2} & Section~\ref{sec:separation} \IEEEauthorrefmark{1} \\ \hline
\multicolumn{1}{|c|}{Broadcast with Side Information} & \cite{Bakshi2008} & Section\ref{sec:duality1} \IEEEauthorrefmark{1} \IEEEauthorrefmark{2} & Section~\ref{sec:duality2} \IEEEauthorrefmark{1}  \\ \hline
\multicolumn{1}{|c|}{Network Ahlswede-Korner} & \cite{Cohen2009}\IEEEauthorrefmark{2} & --- & Section~\ref{sec:ahlswede} \IEEEauthorrefmark{1} \\\hline
\end{tabular}
}
\caption{Existing results and conditions. \IEEEauthorrefmark{1} indicates constraints on the sources structure, and \IEEEauthorrefmark{2} indicates constraints on the network topology.}\label{tab:comp}
\end{table*}

\section{Background and Notations}

Throughout, we denote finite random variables with capital letters, \emph{e.g.}, $X$. We assume takes values in the finite set $\mathcal{X}$. By $\mathcal{T}_\epsilon^n(p_X)$ we denote the set of $\epsilon$-strongly typical sequences, with respect to distribution $p_X$.\\
We will represent networks in the following way: let $\mathcal{G} = (\mathcal{V},\mathcal{E},\mathcal{C})$ be an acyclic directed graph with edge set $\mathcal{E}$, vertices $\mathcal{V}$ and an integer valued function on each edge, $\mathcal{C} : E \rightarrow \mathbb{N}$. The value $C(e)$ for $e \in \mathcal{E}$ represents the capacity of the communication link $e$ in bits per unit of time. In addition, let $\mathcal{S} = \{s_1,s_2,\ldots, s_l\} \subset \mathcal{V}$ and $\mathcal{T} = \{ t_1, t_2, \ldots, t_m\} \subset \mathcal{V}$ be a set of $l$ senders, and a set of $m$ terminals respectively; to avoid trivial cases, we also assume that $\mathcal{S} \cap \mathcal{V} = \emptyset$. We will denote values of min-cuts in the network by $\rho(.;.)$. For example, $\rho(s_1;t_1)$ represents the value of the min-cut from $s_1$ to $t_1$, and $\rho(s_1, s_1;t_2)$ the value of the min-cut from both $s_1$ and $s_2$ to $t_2$. Along with the min-cuts, we will also use \emph{the capacity function of the network}, which is defined as :
\begin{align*}
\rho_G (s_x) &= \min_{t \in T} \rho(s_x;t)\\
\rho_G (s_y) &= \min_{t \in T} \rho(s_x;t)\\
\rho_G (s_x,s_y) &= \min_{t \in T} \rho(s_x,s_y;t).
\end{align*}
We will also consider the $n$-delayed network $\mathcal{G}^n = \{ \mathcal{V},\mathcal{E},\mathcal{C}^n \}$ of $\mathcal{G}$, where $C^{n}(e) = n C(e)$ for all $e \in \mathcal{E}$.
\textbf{G\'acs-K\"orner Common Information:} Before we define the common information, the following will be a useful representation

\begin{definition}
Let $p_{X,Y}$ be the joint probability distribution of $X,Y$. Suppose, without loss of generality that the alphabets $\mathcal{X}$ and $\mathcal{Y}$ are disjoint. We denote by the \emph{bipartite representation} of $p_{X,Y}$, the bipartite graph $\mathcal{B}_{X,Y}$ with the following properties:
\begin{itemize}
\item $|\mathcal{X}| + |\mathcal{Y}|$ nodes indexed $n_a$, with $a \in \mathcal{X} \cup \mathcal{Y}$ 
\item an edge between node $n_{x}$ and $n_{y}$, if and only if $p(X = x)>0$ and $P(Y = y|X=i) > 0$
\end{itemize}
\end{definition}
\textbf{Example: } Let $P_{X_1,X_2}$ be defined as in joint probability table below with its corresponding 2-partite graph representation.

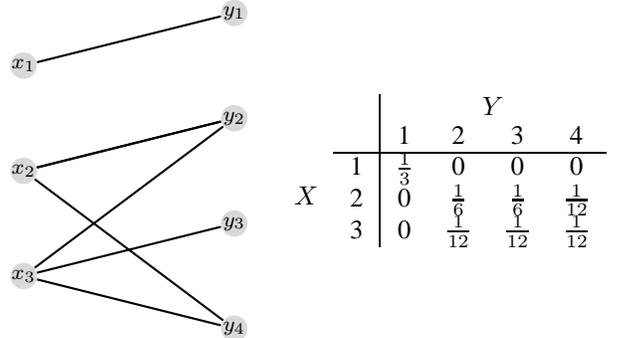
\begin{minipage}{.45\textwidth}
\begin{minipage}[c]{.45\textwidth}
\centering
\begin{tikzpicture}[scale=1.4, auto,swap]
		\foreach \pos/ \name in {{(0,.5)/x_3}, {(0,1.5)/x_2}, {(0,2.5)/x_1}, {(2,0)/y_4},{(2,1)/y_3},{(2,2)/y_2},{(2,3)/y_1}}
        \node[smallvertex] (\name) at \pos {$\name$};
        \foreach  \source/ \dest /\edgename  in {x_1/y_1/1, x_2/y_2/2, x_2/y_2/3, x_2/y_4/4, x_3/y_2/5, x_3/y_3/6, x_3/y_4/7}
        \path[undirect_edge] (\source) -- (\dest);
        \end{tikzpicture}
\end{minipage} 
\begin{minipage}[t]{.45\textwidth}
\centering
\begin{tabular}{c c| cccc}
& & \multicolumn{4}{c}{$Y$} \\
& & 1 & 2 & 3 & 4 \\ \cline{2-6}
\multirow{3}{*}{$X$} & 1 & $\frac{1}{3}$ & 0 & 0 & 0 \\
& 2 & 0 & $\frac{1}{6}$ & $\frac{1}{6}$ & $\frac{1}{12}$ \\
& 3 & 0 & $\frac{1}{12}$ & $\frac{1}{12}$ & $\frac{1}{12}$
\end{tabular}
\end{minipage}
\end{minipage}

\begin{definition}
A set of nodes $\mathcal{C}$ such that there are no outgoing edges from $\mathcal{C}$ is called a connected component of the bipartite graph. We associate with each connected component $\mathcal{C}$ a weight $p(\mathcal{C}) = \sum_{n_x,n_y \in C} P(X=x, Y=y)$. We call \emph{the common information decomposition} of $p_{X,Y}$, the decomposition of the bipartite graph into a maximal number of connected components $\mathcal{C}_1,\ldots,\mathcal{C}_{k}$. Moreover, we denote by the common information $K_{X,Y}$ the random variable representing the index of the connected component generated, with the natural distribution $(p(\mathcal{C}_1),\ldots,p(\mathcal{C}_k))$. The entropy of this random variable is:
\begin{align}
H(K_{X,Y}) = H(\mathbf{\mathcal{C}}) = \sum_{i=1}^k p(\mathcal{C}_i) \log \left( \frac{1}{p(\mathcal{C}_i)}\right) .
\end{align} 
\label{def:common_inf}\end{definition}
Definition of \ref{def:common_inf} is equivalent to the usual definition of common information in the 2 random variables setting given in \cite{Gacs:1973vg} and \cite{Witsenhausen:1975vk}, that is :
\begin{align}
K_{X,Y} = \underset{H(U|X) = H(U|Y) = 0}{\text{argmax}} H(U)
\end{align}

\begin{remark}It is easy to see that the connected components define an equivalent class on $x\in\mathcal{X}$ and on $y \in \mathcal{Y}$. More precisely, we say $x' \in [x]_{X,Y}$ if there is a path between $x$ and $x'$ in the bipartite graph $\mathcal{B}_{X,Y}$. Similarly, we can consider the equivalent class $[y]_{X,Y}$.
\end{remark}

We also introduce the notion of \emph{source decomposition} of $X$ and $Y$. This corresponds to splitting $X$ and $Y$ into three sources $X'$, $Y'$, and $K_{X,Y}$, with the constraint that both $X$ and $Y$ should agree on the value of $K_{X,Y}$.
\begin{definition}
Let $X$ and $Y$ be two sources and $K_{X,Y}$ be their common information. We call source decomposition of $X$ and $Y$, the bijections $X \rightarrow (X',K_{X,Y})$ and $Y \rightarrow (Y',K_{X,Y})$.
\end{definition}
This source decomposition is the main component of our separation scheme. Note that if $K_{X,Y} \neq 0$, it is always possible to construct non-trivial source decompositions of $X$ and $Y$. In fact, there exist a very simple way to construct the random variables $X'$ and $Y'$ given the connected component decomposition of $p_{X,Y}$. 

Let us represent $X$ in two phases, first by expressing in which connected component of the decomposition of $p_{X,Y}$ it is, and then which value among the possible candidates in the connected component it takes. That is, we express the index of the connected $K_{X,Y}$ in which $X$ belong using approximately $H(K_{X,Y})$ bits per symbol, then express $X$ given $K_{X,Y}$. This can be done by giving once again an index. However, as we know the connected component, the index set is limited to those that are in the connected component $C_{i}$. Call this index $X'$.

It is clear that the transformation $X \rightarrow (X',K_{X,Y})$ is a bijection and that a similar construction can be done for $Y$. \\
The following properties follow from the construction of $K_{X,Y}$, $X'$ and $Y'$.

\begin{lemma}(Source Decomposition Inequalities)

\begin{itemize}
\item $H(X'|K_{X,Y}) = H(X|K_{X,Y})$.
\item $H(K_{X,Y}) \leq I(X;Y)$ with equality if and only if $X$ is independent of $Y$ given $K_{X,Y}$.\\
\item $H(X) \geq H(X|K_{X,Y}) \geq H(X|Y)$ and $H(Y) \geq H(Y|K_{X,Y}) \geq H(Y|X)$. \\
\item Let $(X^n,Y^n)$ be $n$ outcomes generated \emph{i.i.d.} from $p_{X,Y}$. Then, the common information of $(X^n,Y^n)$ is given by $K^n_{X,Y} = (K_1,K_2,\ldots,K_n)$ where $K_i$ is the common information between $X_i$ and $Y_i$.
\end{itemize}
\label{prop:source_decomp}\end{lemma}

\noindent \textbf{Proof:} The first property follows from the definition of $X'$. The second property follows from this simple decomposition:
\begin{align}
 I(X;Y) & = I(X,K_{X,Y};Y)\label{eq:bij}\\
& = I(K_{X,Y};Y) + I(X;Y|K_{X,Y}) \label{eq:chain}\\
& = H(K_{X,Y}) + I(X;Y|K_{X,Y}) \label{eq:bij2}.
\end{align}
where (\ref{eq:bij}) follows from the fact that $K_{X,Y}$ is a deterministic function of $X$, (\ref{eq:chain}) from the chain rule for mutual information, and (\ref{eq:bij2}) from the fact that $K_{X,Y}$ is a deterministic function of $Y$.
The third property is a consequence of :
\begin{align}
H(X|Y) &= H(X) - I(X;Y)\\
&= H(X',K_{X,Y}) - I(X;Y) \label{eq:bijection}\\
& \leq H(K_{X,Y}) + H(X|K_{X,Y}) - H(K_{X,Y}) \label{eq:prop1}
\end{align}
where (\ref{eq:bijection}) follows from the fact that $X \rightarrow (X',K_{X,Y})$ is a bijection and (\ref{eq:prop1}) from property 1 and 2, and the chain rule. The proof for $Y'$ is similar. \qed

We also introduce a property for chains of random variables. We will then show that this property results in a data-processing like inequality for common information.

\begin{lemma}\label{lem:equiv_class}
Let $X \leftrightarrow Y \leftrightarrow Z$. For any $x \in \mathcal{X}$ denote by $[x]_{X,Y}$ and $[x]_{X,Z}$ the equivalent class under $p_{X,Y}$, and under $p_{X,Z}$, respectively. Then, for any $x$ such that $p_{X,Z}(x,z) > 0$ for some $z \in \mathcal{Z}$, we have:
\begin{align}
[x]_{X,Y} \subseteq [x]_{X,Z}
\end{align}
\end{lemma}
%
\noindent \textbf{Proof:}
Recall that the bipartite representation of $p_{X,Y}$ creates an equivalence class on $\mathcal{X}$. Namely, we say $x \in [x']_{X,Y}$ if there exist a path in the bipartite graph $\mathcal{B}_{X,Y}$ between $x$ and $x'$. Similarly, the bipartite representation of $p_{X,Z}$ creates an equivalence class $[x]_{X,Z}$.
We want to show that for any $x \in \mathcal{X}$ such that $p(x,z) >0$ for some $z \in \mathcal{Z}$, $x' \in [x]_{X,Y}$ implies that $x' \in [x]_{X,Z}$. 
This means that $[x]_{X,Y} \subseteq [x]_{X,Z}$.

To show this, suppose that there is an $x' \in [x]_{X,Y}$ such that $x' \notin [x]_{X,Z}$. 
Then, it must mean that there is no path between $x$ and $x'$ in $\mathcal{B}_{X,Z}$. 
However, as $p(x,z) >0$ for some $z$, and using $p(x,z) = \sum_y p(x)p(y|x)p(z|y)$, there must be a $y$ such that $p(y|x)>0$ and $p(z|y)>0$. In other words there is a path from $x$ to $y$ in $\mathcal{B}_{X,Y}$. 
In addition, since $x' \in [x]_{X,Y}$ there must be a path from $x'$ to $x$ in $\mathcal{B}_{X,Y}$.
Adding this to the previous fact implies that there must be a path from $x'$ to $y$ in $\mathcal{B}_{X,Y}$. 
Finally, as $p(z|y) > 0$, it means that both $p(x',z)$ and $p(x,z)$ are strictly positive. 
Therefore, there must be a path between $x$ and $x'$ in $\mathcal{B}_{X,Z}$, or, equivalently, that $x' \in [x]_{X,Z}$, which is a contradiction. \qed

The previous lemma has the following immediate corollary:
\begin{corollary}[Data Processing Inequality for Common Information]
Let  $X \leftrightarrow Y \leftrightarrow Z$,  and denote by $K_1$ and $K_2$ the common information between $X$ and $Y$, and between $X$ and $Z$ respectively. Then, $H(K_2|K_1) = 0$, and $H(K_2) \geq H(K_1)$.
\label{lem:data_proc_K} \end{corollary}

\begin{remark}
The previous results can be trivially extended to longer Markov chains $X_1 \leftrightarrow X_2 \leftrightarrow \ldots \leftrightarrow X_m$.
\end{remark}

\begin{remark} \label{remark:nested}
It is easy to see that we can obtain $[x]_{X,Z}$ by taking the union of possibly more than one set $[x]_{X,Y}$. This gives a distinct nested structure to the equivalence classes, where each alphabet is a result of the unions over the previous ones in the chain.
\end{remark}
%
\section{Separation of Source and Network Coding }\label{sec:separation}

\subsection{Fundamental Limits}

Consider a network $G$ with source nodes $s_1,\ldots, s_l$ and destination terminals $\mathcal{T} = \{ t_1,\ldots,t_m\} \subset \mathcal{V}$. At the source node $s_1,\ldots,s_l$, \emph{i.i.d.} copies of $X_1,\ldots,X_l \sim p_{X_1,\ldots,X_n}$ are generated. The multi source multicast network coding problem is transmitting reliably the sources $X_1,\ldots,X_l$ generated at $s_1,\ldots,s_l$, respectively, to all terminals $t \in \mathcal{T}$. A fundamental theorem in network coding, is the min-cut/max-flow theorem \cite{Ahlswede:2000th,Song:2001tf}, stated here only for two source. A general formulation to the $l$-source case can be found in \cite{Song:2001tf}.
\begin{theorem}[Min-cut/max-flow] It is possible to transmit sources $X$ and $Y$ generated at $s_x$ and $s_y$ to all terminals in $t \in \mathcal{T}$, if and only if:
\begin{align}
H(X|Y) &\leq \rho_G(s_x) \nonumber\\
H(Y|X) &\leq \rho_G(s_y) \nonumber\\
H(X,Y) &\leq \rho_G(s_x,s_y).
\end{align}
\label{thm:network_coding}\end{theorem}
If the above conditions are satisfied we say that the network coding problem $\text{NC}(G,p_{X,Y})$ is feasible.

Note that if the sources are independent, the conditions in Thm.~\ref{thm:network_coding} simplify to:
\begin{align}
H(X) &\leq \rho_G(s_x) \nonumber \\
H(Y) &\leq \rho_G(s_y) \nonumber \\
H(X) +H(Y) &\leq \rho_G(s_x,s_y). \label{eq:indep_network_coding}
\end{align}
For independent sources, we use to notation $\text{NC}\left(G,p_X p_Y \right)$ to denote a network coding problem. The generalization of these results to the $m$-source case can be found in \cite{Song:2001tf}.

Proofs of Theorem~\ref{thm:network_coding} rely on either random linear network coding, along with maximum-likelihood or minimum entropy decoding at the receiver (see \cite{ho2004network}), or using traditional random coding and joint typicality (\cite{Song:2001tf}). These methods are high-complexity, and in general, it is not known how to construct efficient-codes to solve the general multisource multicast problem. In the next section, we propose a scheme whose efficiency depends on the value of the G\'acs-K\"orner common information.

\subsection{The case of two sources}
The results on G\'acs-K"orner suggest the following coding scheme for the problem with 2 senders:
\begin{const}(Separation by Source Decomposition)
\begin{itemize}
\item Let $\mathcal{G}^n$ be the $n$-delayed network of $\mathcal{G}$.
\item Consider a sequence of $n$ source outcomes $(x^n,y^n)$, and let $k^n_{X,Y}$ be the sequence of common information random variables. If $k^n \notin T^n_\epsilon(p_{K_{X,Y}})$ encode $k^n$ by its binary representation of size $\lceil n log(|\mathcal{K}|)\rceil$.
\item Else, if the sequence $k^n$ is typical, express $x_i$ using $H(X|K_{X,Y}= k_i)$ for all $0 < i \leq n$. In total to express $x^n$, this necessitates $\sum_{i=1}^n H(X|K_{X,Y} = k_i) = n(H(X|K) + O(\frac{1}{n}))$ bits. Call the encoded sequence $\mathbf{x'}$.
\item Similarly, express $y^n$ using $n(H(Y|K_{X,Y}) + O(\frac{1}{n}))$ bits. Call this encoded sequence $\mathbf{y'}$.
\item Express $k^n$ using $n(H(K_{X,Y}) + O(\frac{1}{n}))$ bits. Call the resulting encoded sequence $\mathbf{k'}$
\item Construct an expanded graph $\mathcal{G'}^n$ where one adds 3 latent sources $X',Y',K_{X,Y}$ generated at nodes $s_{X'}$, $s_{Y'}$ and $s_{K}$. Also add infinite capacity egdes connecting $s_{X'}$ to $s_{X}$, $s_{Y'}$ to $s_{Y}$, and $s_{K}$ to both $s_X$ and $s_Y$.
\item If the rates $(nH(X|K_{X,Y}),nH(Y|K_{X,Y}),nH(K_{X,Y}))$ can be supported by the expanded network $\mathcal{G'}^n$ as if they were independent, transmit $\mathbf{x'}$, $\mathbf{y'}$ and $\mathbf{k'}$ using a simple network coding scheme for independent sources.
\end{itemize}
\end{const}
Let us verify that this scheme indeed yields an achievable solution to the multi source multicast problem. If the rates $(nH(X|K_{X,Y}),nH(Y|K_{X,Y}) ,nH(K_{X,Y}))$ are supported, then all receivers $t_1,\ldots,t_{m}$ receive reliably $\mathbf{k'}$, $\mathbf{x'}$ and $\mathbf{y'}$, which they use to recover $x^n$ and $y^n$. Moreover, since at nodes $s_{X}$ and $s_{Y}$, it is possible to agree on the value of $K_{X,Y}$, all the operations at nodes $s_{Y'}$, $s_{X'}$ and $s_{K}$ can be simulated at nodes $s_X$ and $s_Y$. \emph{Note that this scheme can be implemented with zero-error low complexity codes if one uses low-complexity network codes (say linear codes) for the transmission as encoding of $\mathbf{x'}$, $\mathbf{y'}$ and $\mathbf{k'}$ can be done separately using simple single-source lossless source codes}.

The next theorem characterizes as a function of $H(X|K_{X,Y})$ and $H(Y|K_{X,Y})$ networks than can be decomposed by the encoding above.
\begin{theorem}\label{thm:GK_separation}
Let $(G,C,p_{X,Y})$ be a feasible problem. Then for $n$ sufficiently large, $(G^n,p_{X,Y})$ is feasible by source decomposition if and only if:
\begin{align}\label{eq:GK_cond}
&H(X|K_{X,Y}) \leq \rho_G(s_x) \nonumber \\
&H(Y|K_{X,Y}) \leq \rho_G(s_y) \nonumber \\
H(X|K_{X,Y}) + &H(Y|K_{X,Y}) + H(K_{X,Y}) \leq \rho_G(s_x,s_y)
\end{align}
\end{theorem}

Before we proceed to the proof, let us discuss the sufficient condition of Thm.~\ref{thm:GK_separation}. First of all, since the problem $(G,p_{X,Y})$ is feasible, we have by Thm.~\ref{thm:network_coding}:
\begin{align}
H(X|Y) &\leq \rho_G(s_x)\nonumber\\
H(Y|X) &\leq \rho_G(s_y)\nonumber\\
H(X,Y) &\leq \rho_G(s_x,s_y). \label{eq:cond1}
\end{align}
Next, if the conditions:
\begin{align}
H(X) &\leq \rho_G(s_x)\nonumber\\
H(Y) &\leq \rho_G(s_y)\nonumber\\
H(X)+H(Y) &\leq \rho_G(s_x,s_y) \label{eq:cond2}
\end{align}
are satisfied, then there is no need for complex distributed source coding, and we can simply compress each source separately, and send them as if they were independent. However, since :
\begin{align}
& H(X|K_{X,Y}) \leq H(X) \nonumber\\
& H(Y|K_{X,Y}) \leq H(Y) \nonumber\\
H(X|K_{X,Y}) + & H(Y|K_{X,Y}) + H(K_{X,Y}) \leq H(X)+H(Y)
\end{align}
the conditions of Thm.~\ref{thm:GK_separation} are weaker than (\ref{eq:cond2}), allowing us to send independent sources (after decomposition) to a much larger set of networks. They are, however, stronger than (\ref{eq:cond1}), meaning that there exist feasible network coding problems that are not separable by the suggested source decomposition, yet are solvable with joint source-network codes.\\

\noindent \textbf{Proof:} Consider the expanded network $\tilde{G}$ where we have added nodes $s_{X'}$, $s_{Y'}$ and $s_{K}$, and a feasible problem $(G,C,p_{X,Y})$. By \thmref{thm:network_coding}, we have that:
\begin{align}
H(X|Y) &\leq \rho_G(s_x)\nonumber\\
H(Y|X) & \leq \rho_G(s_y)\nonumber\\
H(X,Y) &\leq \rho_G(s_x,s_y). \label{eq:netcoding}
\end{align}
Further, as the edges between the $s_X, \, s_Y$ and the latent-sources have infinite capacity, it is easy to verify that, on the expanded network $\tilde{G}$, we have :
\begin{align}
\rho_{\tilde{G}}(s_{X'}) &= \rho_G(s_X)\nonumber\\
\rho_{\tilde{G}}(s_{Y'}) &= \rho_G(s_Y)\nonumber\\
\rho_{\tilde{G}}(s_{K}) &= \rho_G(s_X,s_Y)\nonumber\\
\rho_{\tilde{G}}(s_{X'},s_{Y'}) &= \rho_G(s_X,s_Y)\nonumber\\
\rho_{\tilde{G}}(s_{X'},s_{K}) &= \rho_G(s_X,s_Y)\nonumber\\
\rho_{\tilde{G}}(s_{Y'},s_{K}) &= \rho_G(s_X,s_Y)\nonumber\\
\rho_{\tilde{G}}(s_{X'},s_{Y'},s_K) &= \rho_G(s_X,s_Y).
\end{align}
Using Lemma~\ref{prop:source_decomp} and (\ref{eq:netcoding}) we get:
\begin{align}
H(K_{X,Y}) &\leq \rho_{\tilde{G}}(s_{K}) \nonumber \\
H(X|K_{X,Y})+H(K_{X,Y}) &\leq \rho_{\tilde{G}}(s_{X'},s_{K}) \nonumber\\
H(Y|K_{X,Y}) + H(K_{X,Y}) &\leq \rho_{\tilde{G}}(s_{Y'},s_{K}) \nonumber \\
\end{align}
Therefore, $\left(\tilde{G}^n,(nH(X|K_{X,Y}),H(Y|K_{X,Y}),H(K_{X,Y}))\right)$ is feasible if and only if:
\begin{align}
H(X|K_{X,Y}) \leq \rho_{\tilde{G}}(s_X) \nonumber \\
H(Y|K_{X,Y}) \leq \rho_{\tilde{G}}(s_Y) \nonumber \\
H(X|K_{X,Y}) +H(Y|K_{X,Y}) &\leq \rho_{\tilde{G}}(s_{X'},s_{Y'}) \label{eq:redundant}\\
H(X|K_{X,Y}) + H(Y|K_{X,Y}) +H(K_{X,Y}) &\leq \rho_{\tilde{G}}(s_{X'},s_{Y'},s_K) \label{eq:not_redundant}
\end{align}
Finally, it is easy to see that (\ref{eq:not_redundant}) implies (\ref{eq:redundant}), so the latter can be dropped.\qed

The following corollary is immediate, and shows that for if the common information is large enough, any feasible multisource multicast problem can be solved by source decomposition.

\begin{corollary}\label{cor:source_dec_opt}
If $H(K_{X,Y}) = I(X;Y)$, then any feasible $(G,p_{X,Y})$ is feasible by source decomposition.
\end{corollary}
\noindent \textbf{Proof:} If $H(K_{X,Y}) = I(X;Y)$, then it must be the case that $H(X') = H(X|Y)$, and $H(Y') = H(Y|X)$. Using \thmref{thm:GK_separation} yield directly the desired result. \qed

\subsection{From $2$ sources to $m$ sources}

Let $\{ X_{1,i},X_{2,i},\ldots X_{l,i}\}_{i=1}^\infty$ be $l$ random \textit{i.i.d.} processes in $\mathcal{X}_1 \times \mathcal{X}_2 \ldots \times \mathcal{X}_l$ from joint distribution $p_{X_1,X_2,\ldots,X_l}$. Suppose without loss of generality that $\mathcal{X}_1 \cap \mathcal{X}_2 \ldots \cap \mathcal{X}_l = \emptyset$. 

Denote by the \emph{$m-$partite representation of $p$} the m-partite graph such that there exist an hyperedge between $x_1,\ldots,x_m$ if $p(x_1,\ldots,x_m) > 0$. Similarly, let $K$ be the index of the maximum disjoint component decomposition of the $m$-partite graph. Then we have the following coding scheme:

\begin{const}(Source Decomposition for $l$ sources)
\begin{itemize}
\item Let $(k_1,\ldots,k_n)$ be sequences of common information variables. Then, we can encode $(k_1,\ldots,k_n)$ using about $nH(K)$ bits.
\item Encode $(x_{j,1},\ldots,x_{j,n})$ using about $nH(X_j|K)$ bits.
\item if the network can support $(H(X_1|K),\ldots,H(X_m|K),H(K))$ we have a simple coding scheme (send data as if independent).
\end{itemize}
\end{const}

The next theorem is a generalization of Theorem~\ref{thm:GK_separation}, and follows from essentially the same steps:
\begin{theorem}\label{thm:lsources}
Let $G$ be a network, and $(G,p_{X_1,\ldots,X_l})$ be a feasible multicast problem. Then, it is feasible by network decomposition if:
\begin{align}
& H(X_j|U) \leq \rho_G(s_j) \quad j = 1 \ldots m \nonumber \\
& \sum_{j=1}^m H(X_j|U) + H(U) \leq \rho_G(s_1,\ldots,s_j)
\end{align}
\end{theorem}

\begin{remark}
Note that $U$ does not capture entirely the common information in the sources. In particular, it only looks at common information \emph{shared by all the sources together}, without looking at say pairwise common information. Therefore, the result above is useful if the common information over all variables is large enough. In other cases, one might attempt to improve on Theorem~\ref{thm:lsources} by considering \emph{ordered} sets of variables and their common information, though such a scheme might have a significant computational burden.
\end{remark}

\subsection{Comparison with other notions of Separation}
Other notions of separations are introduced in \cite{Ramamoorthy:2006vr} and \cite{Han:2010wa}. In both these works, conditions are given in terms of the network topology for the separation to hold--- in \cite{Han:2010wa} in terms of polymatroidal conditions of the min-cuts, and in \cite{Ramamoorthy:2006vr} in terms of existence of some flows in the network. Interestingly, these conditions are orthogonal to the conditions for separation by Source Decomposition. Therefore, it is not surprising that you can find instances where Separation by Source Decomposition is possible, while the separation schemes in  \cite{Ramamoorthy:2006vr} and  \cite{Han:2010wa} fail. We illustrate one such examples next, by considering a simple source correlation. Namely, we let $X$ and $Y$ be generated at nodes $s_x$ and $s_y$, respectively, and we let $H(X) = H(Y) = 2$, and $H(X,Y) = 3$. Furthermore, we suppose $X$ and $Y$ have the following form:
\begin{align}
X & = (X',K_{X,Y}) \nonumber \\
Y & = (Y',K_{X,Y})
\end{align}
such that such that $H(K_{X,Y}) = I(X;Y) = 1$. $(X,Y)$ then meets the conditions of Corollary~\ref{cor:source_dec_opt}. We then consider the network in Fig.~\ref{fig:counter_example} with unit capacity edges, which is the counter-example provided in \cite{Ramamoorthy:2006vr}, and show a simple source decomposition which solve this problem.

\begin{figure}
\centering
	\begin{tikzpicture}[scale=1.15, auto,swap]
		\foreach \pos/ \name in {{(2,0)/a},{(3,1)/b},{(3,2)/c},{(3,3)/d},{(2,4)/e},{(6,0)/f},{(6,4)/g},{(5,2)/h},{(5,5)/j},{(5,-1)/i},{(7,2)/k},{(0,0)/l},{(0,2)/m},{(0,4)/n}}	
		 \node[vertex] (\name) at \pos {$\name$};
		 \foreach \source/ \dest /\edgename in {a/i/X',a/b/X'+K,b/c/X'+K,d/c/Y'+K,e/d/Y'+K,e/j/Y',d/g/Y'+K,b/f/X'+K,c/h/X'+Y'+2K,h/g/X'+Y'+2K,h/f/X'+Y'+2K,h/k/X'+Y'+2K,i/f/X',j/g/Y',l/a/X',m/a/K,m/e/K,n/e/Y'}
      		 \path[edge] (\source) -- node[weight,sloped,above] {$\edgename$} (\dest);
      		 \path[edge] (i) -| node[weight,below] {$X'$} (k);
      		 \path[edge] (j) -| node[weight,above] {$Y'$} (k);
		\foreach \vertex/ \name in {{a/s_x},{e/s_y}}
       		 \node[source] at (\vertex) {$\name$};
       	\foreach \vertex/ \name in {{f/t_1},{g/t_2},{k/t_3}}
       		 \node[receiver] at (\vertex) {$\name$};
        \foreach \vertex / \name in {{l/X'},{n/Y'},{m/K}}
        		\node[supersource,minimum size=20pt] at (\vertex) {$\name$};		 
     \end{tikzpicture}
       		\caption{Separation by source decomposition for the counter example in \cite{Ramamoorthy:2006vr}}\label{fig:counter_example}
\end{figure}
\section{Broadcast with Side-Information}

\subsection{Fundamental Limits}

In this section, we study a network broadcast problem where the destinations have side information. Again, while the problem is considered solved when there are no limitation on the complexity, when considering efficient solutions we explore the duality between this problem and the degraded message set broadcast. 

Consider a network $\mathcal{G}$, with single source $ s \in \mathcal{V}$, and destinations $\mathcal{T} = \{ t_1, \ldots, t_m\} \subset \mathcal{V}$. The source node $s$ generates \emph{i.i.d} copies of $X \sim p_X$. The destination node $t_i$, $1\leq i \leq m$ has access to side information $Y_i$, and suppose without loss of generality that the decoders are ordered such that $H(X|Y_1) \leq H(X|Y_2) \ldots \leq H(X|Y_m)$. The goal of the network broadcastis to transmit reliably the source $X$ from node $s$ to all destination terminals $t_1,\ldots,t_m$. The following Theorem characterizes the sufficient and necessary condition for feasibility of a network broadcast problem with side-information,

\begin{theorem}[\cite{Bakshi2008}]\label{thm:broad_side_inf}
It is possible the transmit source $X$ generated at $s$ to all terminals $t \in \mathcal{T}$, if and only if:
\begin{align}
H(X|Y_i) \leq \rho(s;t_i) \quad 1 \leq i \leq m 
\end{align}
If the above conditions are satisfied, we say the broadcast problem $\text{BSI}(G,p_{X,Y_1,\ldots,Y_m})$ is feasible.
\end{theorem}
%

The proof of Theorem~\ref{thm:broad_side_inf} relies on random linear network coding, and joint typicality decoding at the receivers. Indeed, the cut constraints guarantee that sufficienly many degrees of freedom are received at the terminals, which lead to a successful joint typicality decoding. As in network coding with correlated sources, this scheme leads to high complexity.
Another approach would consist in using efficient codes at the sources in a separation approach. However, the structure of the code may be broken by the linear operations in the network. Indeed, the cut constraints only guarantee that suffiently many bits will arrive at the destination. Instead, we will impose stronger conditions on the network to derive efficient code constructions.

\subsection{Degraded Message Set and Low-complexity Coding}\label{sec:duality1}

To achieve lower complexity, we would like to use structured codes. However, as we pointed out before, the structure of the code may not be preserved by the network operations. Therefore, we relate the problem to the problem of degraded message set broadcast, which we recall below:
\begin{definition}\label{def:degraded_message_set}
Let $\mathcal{G}$ be a network, and consider a single source $s \in \mathcal{V}$, and destinations $\mathcal{T} = \{ t_1,\ldots, t_m\}$. The source $s$ generates $m$ independent random variables $\{ M_1,\ldots,M_n\}$. We say a degraded message set broadcast problem $\text{DMB}(\mathcal{G},(M_1,\ldots,M_m))$ is feasible if there exist linear encoding and decoding functions on edges such that any terminal $t_i$ receives reliably $M_1,\ldots,M_i$.
\end{definition}
The problem of degraded message set broadcast has been studied extensively in the literature, see \cite{korner1977general, prabhakaran2007, Koetter:2003vx}. In general, the capacity region of the degraded message set broadcast is unknown, however there exist simple instances that are fully solved. For example, when the number of terminals is 2, the problem of degraded message set broadcast can be cast as a simple multicast problem, for which there exist simple linear codes \cite{Koetter:2003vx}. An interesting relationship between this problem and the problem of \emph{low-complexity} broadcast with side information is formulated below:
\begin{theorem}\label{thm:duality_degraded}
Consider a network broadcast with side information problem $(\mathcal{G},X,(Y_1,\ldots,Y_m))$, and let $X \rightarrow Y_1 \rightarrow Y_2 \rightarrow \ldots \rightarrow Y_m$ form a Markov Chain. If the degraded message set broadcast $(\mathcal{G},(M_1,\ldots,M_m))$ --- from the same source and to the same destinations--- with $H(M_1,\ldots, M_i) = H(X|Y_i)$ is feasible, then there exist \emph{error-correcting} efficient codes for the network broadcast with side information problem.
\end{theorem}
\noindent \textbf{Proof Sketch:}
The proof relies on so-called \emph{successive description codes} such as Monotone Chain polar codes \cite{Salamatian2015}. These Slepian-Wolf codes have the property that some subsets of the bits are useful in decoding at both rates $(R_1,R_2)$ and rates $(\tilde{R}_1,\tilde{R}_2)$, for $n$ large enough. 

Let $h_i = H(X|Y_i)$, an arbitrary $\epsilon > 0$, and consider a large block-length $n$. 
In our context, by the data-processing inequality on the Markov Chain, we have $h_1 \leq h_2 \leq \ldots \leq h_m$. 
Denote by $\bG_i$ the generator matrix of size $n(h_i + \epsilon) \times n$ that encodes $X^n$ into $Z_i^{n(h_i + \epsilon)}$, \emph{i.e.} $\bZ_i = \bG_i \bX$.
The decoder $i$ then uses $\bZ_i$ and $\bY_i = Y_i^n$ to recover $\bX$.
The successive description property guarantees that for any $i,j \in \{1,\ldots,m\}$ with $i<j$, there are good encoding matrix $G_i$, such that the rows of $G_i$ are a subset of the rows of $G_j$. 
By rearranging the rows of $G_j$, this implies that $\bZ_i$ can be written as $\bZ_i = [\bZ_j, \bZ_{i \backslash j}]^T$.
The rest of the proof follows directly by using degraded message set broadcasting, to transmit the messages $M_i = \bZ_i$. \qed

Theorem~\ref{thm:duality_degraded} shows a duality between a low-complexity approach to broadcast with side-Information, and the problem of degraded message set broadcast. As in the previous section, we can also use a scheme based on the common information to further simplify the encoding and decoding scheme, at the expense of stronger constraints on the source structure.

\begin{remark}
The constraint on the strict degradation of the messages can be relaxed by considering other structured codes such as LDPC codes under the Wyner scheme. Indeed, contrary to Monotone-Chain Polar Codes, LDPC do not require a specific set of bits to be sent, as with high probability, any subset of the rows of a randomly generated low density parity check will have good properties.
\end{remark}

\subsection{Using the Common-information}\label{sec:duality2}

Consider the case where there is some common information between the pairs $(X,Y_i)$, and denote it by $K_i$. In this case, we can once again use the structure of the correlation to simplify the coding scheme. The scheme can be explained in few words --- send only $H(X|K_i)$ to the terminal $i$, as it can compute $K_i$ with no help from the senders. Using Lemma~\ref{lem:data_proc_K}, we can guarantee that the description of the $K_i$ are properly nested, and use this to construct a degraded message set broadcast problem. 
The overall coding scheme can be roughly described in the following steps:
\begin{const} (Broadcast with Side-Information using Common Information)
\begin{itemize}
\item Let $n$ be the blocklength, and consider the $n$-delayed network $\mathcal{G}^n$
\item Consider a sequence of $n$ outcomes $(x^n,y_1^n,\ldots,y_m^n)$, and let $(k_1^n,\ldots,k_m^n)$ be the sequence of common information.
\item Otherwise, express $x^n$ using approximately $n(H(X|K_1) + O\left(\frac{1}{n}\right))$ bits by losslessly encoding the sequence of length $n$ of its index in the equivalence set $[x]_{X,Y_1}$. Call this $M_1$.
\item Encode using $n(H(K_1|K_2))$ the index of $[x]_{X,Y_1}$ among the equivalent classes that compose $[x]_{X,Y_2}$, \emph{c.f.} Remark~\ref{remark:nested}, and let this be $M_2$
\item Similarly, express the index of $[x]_{X,Y_i}$ given $[x]_{X,Y_{i-1}}$ for $2 \leq i \leq m$, and let this be $M_i$.
\item Send the messages $M_1,\ldots,M_m$ using degraded message set broadcast.
\end{itemize}
\end{const}
The achievable rates of the above construction are given in the following Theorem:

\begin{theorem}\label{thm:duality_common_inf}
Consider a network broadcast with side information problem $\text{BSI}(\mathcal{G},p_{X,Y_1,\ldots,Y_m})$, and let $X \rightarrow Y_1 \rightarrow Y_2 \rightarrow \ldots \rightarrow Y_m$ form a Markov Chain. If the degraded message set broadcast $\text{DMB}(\mathcal{G},(M_1,\ldots,M_m))$, with $H(M_1,\ldots, M_i) = H(X|K_i)$ is feasible, then there exist a zero-error low-complexity scheme for the network broadcast with side information problem.
\end{theorem}

\noindent \textbf{Proof: }
The proof follows by observing that:
\begin{align}
H(M_1,\ldots,M_i) & = H(X|K_1) + \ldots + H(K_{i-1}|K_{i}) \\
& = H(X|K_1) + \ldots + I(X;K_{i-1}|K_{i}) \\
& = H(X|K_i)
\end{align}
where the last step follows by expanding $I(X;K_j|K_{j-1}) = H(X|K_j) - H(X|K_{j-1})$ and telescoping. \qed

The results of Theorem~\ref{thm:broad_side_inf}, \ref{thm:duality_degraded} and \ref{thm:duality_common_inf} can be seen as three levels of coding complexity for solving the problem of broadcast with side information. By adding a constraint on degradation of the sources and on the degraded message set, we can move from a high-complexity coding, to an error-correcting complexity. Additional structure in the form of Gacs-Korner common information allows for a zero-error efficient encoding.

\section{Ahlswede-Korner Problem in a Network}\label{sec:ahlswede}

We now move to the general Ahswelde-Korner problem. Again, we consider a network $\mathcal{G}$ with two source nodes $s_x$ and $s_y$, and terminals $\mathcal{T} = \{t_1,\ldots,t_m\} \subset \mathcal{V}$. Sources $s_x$ and $s_y$ generate \emph{i.i.d.} copies of $X$ and $Y$, respectively, where $(X,Y) \sim p_{X,Y}$. Contrary to the multicast with correlated sources setting, the goal here is to reliably transmit only $X$. The correlation between $X$ and $Y$ allows the source node $s_y$ to act as a helper. This problem of source-coding with coded side-information has been studied in \cite{ahlswede1975source, marco2009lossless}, and its fundamental limits are known in simple settings where the network is composed of only the 3 nodes.

The general network setting of the Ahswelde-Korner problem has been studied in \cite{Cohen2009}, where lower and upper bounds were given for the rate region. These bounds have been found to be tight for special networks, namely, networks in which there is no mixing between the source $X$ and the source $Y$. In the sequel, we give a simple zero-error low-complexity construction for this problem for general networks, using, once again, common information. The code construction can be explained below:
\begin{const}(Ahlswede-K\"orner using Common Information)
\begin{itemize}
\item Let $\mathcal{G}^n$ be the $n$-delayed network of $\mathcal{G}$.
\item Consider a sequence of $n$ source outcomes $(x^n,y^n)$, and let $k^n_{X,Y}$ be the sequence of common information random variable. If $k^n \notin T^n_\epsilon(p_{K_{X,Y}})$ encode $k^n$ by its binary representation of size $\lceil log(|\mathcal{K}|)\rceil$.
\item Else, if the sequence $k^n$ is typical, express $x_i$ using $H(X|K_{X,Y}= k_i)$ for all $0 < i \leq n$. In total to express $x^n$, this necessitates $\sum_{i=1}^n H(X|K_{X,Y} = k_i) = n(H(X|K) + O(\frac{1}{n}))$ bits. Call the encoded sequence $\mathbf{x'}$.
\item Express $k^n$ using $n(H(K_{X,Y}) + O(\frac{1}{n}))$ bits. Call the resulting encoded sequence $\mathbf{k'}$.
\item If the rates $nH(X|K_{X,Y})$ and $nH(K_{X,Y})$ can be supported by the network from $s_x$ and $s_y$, respectively, transmit $\mathbf{x'}$ and $\mathbf{k'}$ using linear network codes.
\end{itemize}
\end{const}
It is clear that the above scheme is a zero-error low-complexity scheme. The rates that this code construction can achieve are given in the following theorem:

\begin{theorem}[Ahlswede-Korner with Common-Informations]
Let $K_{X,Y}$ be the common information between $X$ and $Y$, and let $A \sim \text{Ber}(\gamma)$ for some $0 \leq \gamma \leq 1$ be independent of $X$ and $Y$. Define $U = A \times K_{X,Y}$. Then, the following conditions are sufficient for reliable communication of $X$ at the terminals $\mathcal{T}$, for any $\gamma$:
\begin{align}
& \rho_G(s_x;t) \geq H(X|U) \nonumber \\
& \rho_G(s_y;t) \geq H(U). \label{eq:AK_bound}
\end{align}
In addition, this can be done in low-complexity.
\end{theorem}

\noindent \textbf{Proof:}
For $\gamma = 0$, this clearly holds, as we are sending the complete representation of $X$ to the terminal. For $\gamma = 1$, we obtain precisely the code construction above, which can be analyzed in a similar way as the source decomposition method for multicast. For $0 < \gamma < 1$, this holds by invoking a simple time-sharing argument. \qed

\begin{remark}
Note that the above construction does not, in general, capture the full Ahlswede-K\"orner region.
Indeed, the classical Ahlswede-K\"orner bound has a similar form as the one given in (\ref{eq:AK_bound}), with the difference that the auxiliary random variable $U$ is chosen among all random variables that satisfy $U \leftrightarrow X \leftrightarrow Y$, and that the bound on $s_y$ is $I(Y;U)$.
To achieve low-complexity, we pick $U = A \times K_{X,Y}$, which verifies the Markov chain condition above, but is not sufficient to represent all the possible random variables.
Note that in this case, the term $I(Y;U)$ can be expressed as $H(U) - h(\gamma)$.
Nevertheless, it is easy to see that (\ref{eq:AK_bound}) better than the Slepian-Wolf region, in which we must decode fully $Y$. Here, we only send at most $nH(K_{X,Y})$ bits from the source $s_y$.
\end{remark}

\section{Final Remarks}

In this work, we have introduced G\'acs-K\"orner common information as a new tool for the design of efficient multisource network problems with correlated sources. 
Through the study of three problems --- multisource multicast, broadcast with side information, and lossless encoding with coded side information --- we showed that the structure of the source correlation can help in designing efficient and powerful codes.
A simple feature of these schemes is the simple coding which does not necessitate a traditional distributed coding approach as in classical Slepian-Wolf problems. Instead, we avoid redundancy by separating the common core of the sources, and by expressing the sources given this common information. This distributed coding scheme, which may be sub-optimal in terms of rates as defined in the Slepian-Wolf problem, still allows us to reduce complex network problems, to simpler instances where sources are set to be independent.
The performance of this separation scheme relies on the existence of a large common information between the variables, in contrast with existing conditions in the literature on the network topology.
\section{Acknowledgements}
The authors would like to thank Emre Telatar for his guidance and input on a first version of this manuscript, in particular for his comments on the separation of source and network coding.

\bibliographystyle{./biblio/IEEEtran}
\bibliography{./biblio/IEEEabrv,references}

\end{document}